\documentclass[letterpaper, 10 pt, conference]{ieeeconf}  

\IEEEoverridecommandlockouts                              
\usepackage{graphicx}
\usepackage{subcaption}
\usepackage{amsmath} 
\usepackage{amssymb}  
\usepackage[ruled,vlined,linesnumbered]{algorithm2e}
\usepackage{amsthm}
\usepackage{xcolor}
\usepackage{cite}
\usepackage{hyperref}
\usepackage{enumerate}

\newtheorem{theorem}{Theorem}
\newtheorem{definition}{Definition}
\newtheorem{problem}{Problem}

\newtheorem{assumption}{Assumption}
\newtheorem{remark}{Remark}
\newtheorem{lemma}{Lemma}
\newtheorem{example}{Example}

\DeclareMathOperator*{\argmax}{arg\,max}

\title{\LARGE \bf
Bayesian Diagnosability and Active Fault Identification

}

\author{Chun-Wei Kong, Jay McMahon, Morteza Lahijanian
\thanks{This work was supported by the Air Force Research Laboratory/RVSW under Contract No. FA945324CX026.}
\thanks{Authors are with the Dept. of Aerospace Engineering Sciences,
        University of Colorado Boulder, USA
        {\tt\small \{firstname.lastname\}@colorado.edu}}%
}

\begin{document}

\maketitle
\thispagestyle{empty}
\pagestyle{empty}

\begin{abstract}
We study fault identification in discrete-time nonlinear systems subject to additive Gaussian white noise. 
We introduce a Bayesian framework that explicitly accounts for unmodeled faults under reasonable assumptions. 
Our approach hinges on a new quantitative diagnosability definition, revealing when passive fault identification (FID) is fundamentally limited by the given control sequence. 
To overcome such limitations, we propose an active FID strategy that designs control inputs for better fault identification. 
Numerical studies on a two-water tank system and a Mars satellite with complex and discontinuous dynamics demonstrate that our method significantly reduces failure rates with shorter identification delays compared to purely passive techniques.

\end{abstract}

\section{Introduction}
Autonomous systems in uncertain or hazardous environments must identify faults quickly and reliably, yet relying on passive inputs (e.g., from default controller) can fail to reveal subtle or latent faults.
To address this, active fault identification (FID) designs diagnostic control inputs to amplify fault signatures, such as lightly tapping brakes, to expose a fault hidden under normal conditions \cite{campbell2015auxiliary}. 
Despite significant advancements, a major challenge remains in FID: the lack of a broadly applicable, quantitative definition of diagnosability---a critical element for understanding FID reliability and the limits of active control, as noted by \cite{heirung2019input} and references therein. 
This work tackles this challenge by proposing a Bayesian framework for both passive and active FID along with a well-defined notion of diagnosability.

Diagnosability in abstract systems, such as graphs or finite automata, has been studied in \cite{maheshwari1976models,yang2007efficient,chwa1981fault}, particularly for analyzing faulty components in multiprocessor computer systems. However, these approaches typically neglect the effects of control inputs and measurement noise, which are critical in dynamic, real-world systems.
Optimal filtering in \cite[Ch. 10]{anderson2005optimal} presents a distinguishability criterion for fault identification under ergodicity and an infinite time horizon but does not address the influence of control inputs.
A recent definition in \cite{vignolles2020overview} incorporates control authority but remains qualitative, lacking an explicit quantitative measure of diagnosability.
Consequently, these studies do not fully capture the complex interplay between fault models, uncertainties, and the influence of diagnostic controls.

Existing model-based active FID methods perform well when the true fault is among a finite set of models.
For instance, \cite{paulson2014guaranteed} ensures output separation for a nonlinear two-tank system with bounded noise, while \cite{mesbah2014active} uses Hellinger distances to distinguish discrete-time fault models under parameter uncertainties. 
This concept was extended in \cite{paulson2017input,martin2018active,palmer2018active,palmer2019analysis} for broader types of faults and uncertainties.
Some approaches combine Kalman filters with reinforcement learning \cite{vskach2017input} or cast fault identification as a partially observable Markov decision process \cite{ragan2024online}. While effective when the true fault is modeled, their performance and guarantees remain unclear otherwise.

We bridge these gaps by introducing a quantitative diagnosability definition for discrete-time nonlinear systems with additive Gaussian white noise. 
Our definition explicitly incorporates control inputs, measurement noise, and potentially unmodeled faults---factors often overlooked in prior work. 
Based on this, we propose a Bayesian-based fault identification framework that provides reliability guarantees even when the true fault lies outside the modeled set.
We show that with sufficiently exciting control sequences, our proposed passive approach achieves asymptotic reliability---even with unmodeled faults---as more data become available.
Since the passive approach may be fundamentally limited 
under a given control sequence, we introduce an active fault identification strategy that selects inputs to maximize a proxy of the separation distance in our diagnosability definition.

In short, our contributions are:
\begin{enumerate}[i.]
    \item A diagnosability criterion tying the selection of fault models and control sequences, 
    \item A passive fault identification algorithm that ensures asymptotic reliability under certain conditions, addressing limitations from unmodeled dynamics,  
    \item An active approach to design controls meeting the necessary condition for diagnosability, and
    \item Comprehensive numerical studies demonstrating the advantages of our methods.
\end{enumerate}


\section{Problem Formulation}
The works aims to develop a framework for accurate fault identification in controlled stochastic systems.
Let  $\mathcal{H}$ denote the set of all possible faults a system can experience, which can be \emph{uncountable} and \emph{unknown}.
Under each fault $h \in \mathcal{H}$, system dynamics and measurement model are given by:
\begin{subequations}
    \label{eq:sys_dyn}
    \begin{align}
        &x_{k+1} = F_h(x_k, u_k) + w_k, &&w_k \sim \mathcal{N}(0,Q), \\ 
        &y_k = G(x_k) + v_k, &&v_k \sim \mathcal{N}(0,R),
    \end{align}
\end{subequations}
where $k\in\mathbb{N}^0$ is the time step, $x_k \in \mathbb{R}^{n_x}$ is the state, 
$u_k \in U \subseteq \mathbb{R}^{n_u}$ is the control,
$y_k \in \mathbb{R}^{n_y}$ is the measurement at time $k$. 
Function $F_h:\mathbb{R}^{n_x} \times U \rightarrow \mathbb{R}^{n_x}$ is the 
vector field of the system under fault $h$,
and $G:\mathbb{R}^{n_x} \rightarrow \mathbb{R}^{n_y}$ is the measurement function.
Terms $w_k \in \mathbb{R}^{n_x}$ and $v_k \in \mathbb{R}^{n_y}$ are random variables, representing the process and measurement noise, respectively. 
We assume $w_k \sim \mathcal{N}(0,Q)$ and $v_k \sim \mathcal{N}(0,R)$ are i.i.d. samples from Gaussian distributions with zero mean and positive definite covariance matrices $Q \succ 0$ and $R \succ 0$, respectively.
Furthermore, we consider a finite set of \emph{modeled} fault hypotheses\footnote{Fault hypothesis can be designed based on domain knowledge or past experiences.}, denoted by
\begin{equation}
    \mathcal{M} = \{m_1,\,m_2,\,\dots,m_{|\mathcal{M}|}\},
\end{equation}
where each hypothesis $m \in \mathcal{M}$ represents a model in the form of~\eqref{eq:sys_dyn}.
We assume that, at least one of the modeled hypotheses in $\mathcal{M}$ is a realizable fault, 
the hypotheses are distinct, and each model is observable and controllable.  



\begin{assumption}
The modeled hypotheses set $\mathcal{M}$ satisfies the following properties:
    \begin{enumerate}
        \item The modeled hypotheses in $\mathcal{M}$ are distinct, i.e., $F_{m} \neq F_{m'}$  for all $m \neq m' \in \mathcal{M},$ where $F_{m}$ is the vector field corresponding to hypothesis model $m$.
        \label{assump:DistinctHypotheses}
    
        \item At least one hypothesis in $\mathcal{M}$ is realizable, i.e., $\mathcal{M} \cap \mathcal{H} \neq \emptyset.$\footnote{Note that this does not imply that $\mathcal{M}$ contains the true fault.}
        
        \item The model corresponding to each hypothesis $m \in \mathcal{M}$ is \emph{observable} and (partially) \emph{controllable}.
        \label{assump:ObsCon}
    \end{enumerate}
\end{assumption}
\noindent
Note that these assumptions are \emph{not} restrictive since we usually design $\mathcal{M}$.



Fault identification (FID) determines which (if any) hypothesis in $\mathcal{M}$ matches the true fault from system observations.
We study two variants of this problem: \emph{passive} FID with fixed controls and \emph{active} FID with designed controls. 

\begin{problem}[Passive FID]
\label{prob:PassiveFaultID}
Consider the stochastic system in \eqref{eq:sys_dyn} under a true fault \(h^* \in \mathcal{H}\)  
with an uncertain initial state 
$x_0 \in \mathbb{R}^{n_x}$ drawn from a Gaussian distribution 
\(x_0 \sim \mathcal{N}(\hat{x}_0,\Sigma_0)\) with mean $\hat{x}_0 \in \mathbb{R}^{n_x}$ and covariance matrix $\Sigma_0 \succ 0$.   
Given a control sequence $u_0,\ldots,u_K \in U$, where $K \in \mathbb{N}^0$, the corresponding state measurements $y_0,\ldots,y_K \in \mathbb{R}^{n_y}$, a set of modeled hypothesis $\mathcal{M}$ that satisfy Assumption~\ref{assump:DistinctHypotheses}, identify true fault $h^*$ if it is in $\mathcal{M}$, otherwise return \textsc{null}, i.e., determine
\[
    m^\mathrm{ID} \;\in\; \mathcal{M} \,\cup\, \{\textsc{null}\}
\]
such that
\[
    m^\mathrm{ID} =
    \begin{cases}
        h^* & \text{if } h^* \in \mathcal{M}, \\
        \textsc{null} & \text{otherwise.}
    \end{cases}
\]
\end{problem}

While Problem~\ref{prob:PassiveFaultID} is important, it often lacks diagnosability under the given control sequence. Consider the following example.
\begin{example}\label{ex:1}
    Let $\mathcal{M} = \{h^*, m\}$ include two systems $h^*:x_{k+1} = x_k + u_k + w_k,\; y_k =  x_k + v_k$ and $m: x_{k+1} = x_k + diag([1.0, 0.5]) u_k + w_k,\; y_k = x_k + v_k$, where $x_k \in \mathbb{R}^2$ and $u_k \in U = \mathbb{R}^2$. Although both systems are observable, controllable, and distinct (satisfying Assumption~\ref{assump:DistinctHypotheses}), it is impossible to identify true fault using control sequence $\{u_k = [u_{1,k},0]^T \}_{k \ge 0}$ for any choice of $u_{1,k} \in \mathbb{R}$
    because the filter estimates of $h^*$ and $m$ remain identical.
\end{example}
In other words, passive FID in Problem~\ref{prob:PassiveFaultID} may be fundamentally limited, motivating the second variant where limited control authority is used to actively perform FID.


\begin{problem}[Active FID]
\label{prob:ActiveFaultID}

Consider the setting in Problem~\ref{prob:PassiveFaultID} and a compact set of admissible control $U_a \subseteq U$.  Design control sequence $u^*_0,\ldots,u^*_{K} \in U_a$ that improves identification of $m^\textsc{ID}$.


\end{problem}

We tackle these problems with a Bayesian framework combining Bayesian filters and hypothesis testing. 
Within this framework, we define diagnosiblity and provide theoretical analysis (Sec.~\ref{sec:bayes_fid}), and accordingly design algorithms to solve Problem~\ref{prob:PassiveFaultID} (Sec.~\ref{sec:passive FID}) and Problem~\ref{prob:ActiveFaultID} (Sec.~\ref{sec:active FID}).


\section{Bayesian Fault Identification}\label{sec:bayes_fid}


\subsection{Bayesian Hypothesis Inference}
Inspired by~\cite{Andersson:CIS:2008}, our FID approach leverages Bayesian inference to sequentially update the probabilities of the modeled hypotheses in $\mathcal{M}$.
Let the \textit{belief} $b_k$ be the probability distribution over $\mathcal{M}$ representing the likelihood of each $m \in \mathcal{M}$ being the true fault at time step $k$.
Formally, for every $k \in \mathbb{N}^0$,
\[
b_k(m) \in [0,1]  \quad \forall m\in \mathcal{M} \quad \text{and} \quad \sum_{m \in \mathcal{M}} b_k(m) = 1.
\]
With an abuse of notation, we use $b_k(\mathcal{M}) \in [0,1]^{|\mathcal{M}|}$ to be the vector of beliefs, whose $i$-th element is the belief $b_k(m_i)$ of the $i$-th hypothesis in $\mathcal{M}$.

Given an initial belief $b_0$, the subsequent beliefs are inferred sequentially using the available information (measurements and controls) in the current and previous $N-1$ time steps.
We refer to $N \in \mathbb{N}^+$ as the \textit{moving window} and define \textit{information} at time step $k$ with moving window $N$ as:
%
%
\[
I_k^N := 
\begin{cases}
\{y_k\} \cup \{(u_{i},y_i)\}_{i=k-(N-1)}^{k-1} 
& \text{if } k \ge N-1,\\
\varnothing & \text{if } k < N-1.
\end{cases}
\]
Then, through the Bayes' rule, we can inductively obtain the next belief, i.e., given $b_{k-1}$,
\begin{equation}
    b_k(m \mid I_k^N) = \frac{p(I_k^N \mid m)\, b_{k-1}(m)}{z} \qquad \forall m\in \mathcal{M},
    \label{eq:belief_update}
\end{equation}
where \(z\) is a normalization factor and \(p(I_k^N \mid m)\) denotes the likelihood of \(I_k^N\) under hypothesis \(m\). For simplicity, we denote the belief updated by $N$ moving window information as $b_k^N$.



\subsubsection*{Information Likelihood Computation}
To apply~\eqref{eq:belief_update}, we compute the information likelihood $p(I_k^N \mid m)$ for each $m \in \mathcal{M}$ using the state estimates from the corresponding model.
Since each model in \(\mathcal{M}\) is observable (Assumption~\ref{assump:DistinctHypotheses}), we adopt the standard filtering approach in \cite[Sec. 10.2]{simon2006optimal}.
Specifically, we require the estimator for each \(m \in \mathcal{M}\) to be a Bayesian Gaussian filterr, where state estimates are Gaussian and the filter follows the standard \emph{predict}–\emph{update} procedure.
This framework encompasses widely used filters such as the Extended Kalman Filter (EKF) and Unscented Kalman Filter (UKF). 

For completeness, we briefly review EKF and its consistent definition here but emphasize that our approach is not limited to EKFs. Given a prior state estimate $x_{k-1} \sim \mathcal{N}(\hat{x}_{k-1}, \Sigma_{k-1})$ and system model~\eqref{eq:sys_dyn} for hypothesis $m$, EKF estimates $x_k \sim \mathcal{N}(\hat{x}_k, \Sigma_k \mid m)$ by:
\begin{subequations}\label{eq:ekf}
    \begin{align}
        &\textit{Predict:}\;
        \begin{cases}
            \hat{x}_{k|k-1} = F_{m}(\hat{x}_{k-1}, u_{k-1}),\\
            \Sigma_{k|k-1} = \phi \Sigma_{k-1} \phi^T + Q_{k-1}, \; \phi = \frac{\partial F_{m}}{\partial x} \big|_{\hat{x}_{k-1},u_{k-1}} \\
            \hat{y}_{k|k-1} = G(\hat{x}_{k|k-1}),\\
            S_{k} = H \Sigma_{k|k-1} H^T + R_k, \; H = \frac{\partial G}{\partial x} \big|_{\hat{x}_{k|k-1}} \\
            K_k = \Sigma_{k|k-1} H^T(S_k)^{-1},
        \end{cases}\\
        &\textit{Update:}\;
        \begin{cases}
            e_{y,k} = y_k - \hat{y}_{k|k-1}, \\
            \hat{x}_k = \hat{x}_{k|k-1} + K_k e_{y,k}, \\
            \Sigma_{k} = (I-K_kH)\Sigma_{k|k-1},
        \end{cases}
    \end{align}
\end{subequations}
where 
\(\hat{y}_{k|k-1}\) is the filter's predicted (expected) measurement, $e_{y,k}$ is the \emph{innovation} (measurement residuals), and $S_k$ is the innovation covariance.

\begin{definition}[Consistent Gaussian Filters]
\label{def:consistent_filter}
A Gaussian filter is said to be \emph{consistent} if its innovations are whitened, i.e., 
\begin{equation}
    \label{eq:innovation estimate}
    e_{y,k} := (y_k - \hat{y}_{k|k-1}) \sim \mathcal{N}(0,S_k), 
\end{equation}
where \(\hat{y}_{k|k-1}\) and \(S_k\) are defined above (see \cite[Ch.10]{simon2006optimal}).
\end{definition}

\begin{remark}\label{remark:general_of_consistency}
    In other Gaussian filters (e.g., UKF), the covariance mapping method may differ, i.e., the computation of $\Sigma_{k|k-1}$ of $S_k$ may vary, but the definition of consistency remains the same.
\end{remark}

Without loss of generosity, given the Gaussian filter's innovation $e_{y,k} \sim \mathcal{N}(0,S_k \mid m)$ for hypothesis $m$,
the joint likelihood of observing $I_k^N$ in~\eqref{eq:belief_update} is~\cite[Sec. 10.2]{simon2006optimal}:
\begin{align}
    p(I_k^N \mid m) &= p(y_k,y_{k-1},\dots,y_{k-(N-1)} \mid  m) \nonumber \\
    & = \Pi^{k}_{i=k-(N-1)} \mathcal{N}(0, S_{i} \mid m). \label{eq:likelihood}
\end{align}
This allows us to compute and update the beliefs $b_k(\mathcal{M})$ over hypotheses $\mathcal{M}$ in~\eqref{eq:belief_update}, forming the basis for the following analysis.

\subsection{Diagnosability under Noisy Measurements}
Noise can reduce the signal-to-noise ratio, making reliable information extraction difficult.
We therefore introduce notions of diagnosability and its fundamental limit for the modeled hypotheses set $\mathcal{M}$ to characterize a necessary condition for Bayesian FID.

\begin{definition}[Diagnosability]
    \label{def:diagnos}
    Consider System~\eqref{eq:sys_dyn} under fault $h^*$, a sequence of controls $u_0,\ldots,u_k$, and a set of modeled hypotheses $\mathcal{M}$ that contains $h^*$. Furthermore, for model $m \in \mathcal{M}$, let $\hat{y}_{m,k|k-1}$ and $S_{m,k}$ be the expected measurement and innovation covariance, respectively, given by a consistent Gaussian filter at time $k$.  The \emph{diagnosibility} $\lambda_k^N$ of $\mathcal{M}$ at time step $k$ for a given time window $N\le k+1$ is defined as:
    \begin{align*}
        \lambda_k^N&(\mathcal{M}) = \min_{m \in \mathcal{M} \setminus \{h^*\}} \Big\{ \frac{1}{N} \sum_{i=k-(N-1)}^k \\ 
        &\mathbb{E} \big[ ( \hat{y}_{h^*,i|i-1} - \hat{y}_{m,i|i-1} )^T 
         S_{m,i}^{-1} ( \hat{y}_{h^*,i|i-1} - \hat{y}_{m,i|i-1} ) \big] \Big\}.
    \end{align*}
    The \emph{diagnosibility} of  $\mathcal{M}$ given $N$ is defined as
    \begin{equation*}
        \lambda^N (\mathcal{M}) = \min \{\lambda_k^N(\mathcal{M}) \mid k \geq N-1\}.
    \end{equation*}
\end{definition}

The diagnosability \(\lambda^N(\mathcal{M})\) quantifies the minimum average separation, over an \(N\)-step window, between the predicted measurements of the true fault \(h^*\) and those of any other hypothesis in \(\mathcal{M}\setminus\{h^*\}\). This separation is measured relative to the uncertainty in the filter (via the innovation covariance \(S_{m,i}\)); hence, a larger \(\lambda^N(\mathcal{M})\) indicates that the true fault is consistently distinguishable from incorrect hypotheses, even under noisy conditions. Conversely, if \(\lambda^N(\mathcal{M})=0\), then at least one false hypothesis produces predictions indistinguishable from \(h^*\), fundamentally limiting fault identification.

\begin{definition}[Fundamental Limit of Diagnosability]
    \label{def:fund_limit}
    Consider the setting in Definition~\ref{def:diagnos}. Modeled hypotheses set $\mathcal{M}$ is called \emph{fundamentally limited}
    for control sequence $u_0,u_1,\ldots$ and time window $N$,
    if its diagnosability $\lambda^N (\mathcal{M}) = 0$. 
    Otherwise, it is not fundamentally limited.
\end{definition}

Definition~\ref{def:fund_limit} serves as a necessary condition for reliable fault identification via Bayesian belief update with Gaussian filter estimates. 
In the following, we discuss a sufficient condition on the growth of \(\lambda^N\) to guarantee asymptotic diagnosability. For this analysis, we recall the notion of filter stability, which is related to the observability of the underlying nonlinear systems \cite[Theorem 5.2]{song1992extended}.


\begin{definition}[Stable Gaussian Filter]
\label{def:stable_filter}
    A Gaussian filter is stable if
    $\lim_{k \rightarrow \infty} \mathbb{E}[y_k - \hat{y}_{k|k-1}]$ is bounded and 
    $\lim_{k \rightarrow \infty} S_k = S_{ss} \succ 0.$
\end{definition}

We now state the sufficient condition on $\lambda^N(\mathcal{M})$ to ensure asymptotic fault diagnosability. 
Proofs of Theorem~\ref{theorem:diagnosability} and all other technical results are provided in the arXiv version.

\begin{theorem}[Asymptotic fault diagnosability]
\label{theorem:diagnosability}
Consider the following conditions:
\begin{enumerate}
    \item The filter of each hypothesis \( m \in \mathcal{M} \) is designed to be consistent to its corresponding dynamics.
    \item The true fault \( h^* \in \mathcal{M} \).
    \item The control sequence $\{u_k\}_{k \geq 0}$ ensures that $\mathcal{M}$ is not fundamentally limited and all filters are stable. 
    \item The growth of diagnosibility metric \( \lambda^N(\mathcal{M}) \) with respect to $N$ is asymptotically unbounded.
\end{enumerate}
If the above conditions hold, then the Bayesian belief via~\eqref{eq:belief_update} and~\eqref{eq:likelihood} asymptotically converges to 1, i.e., 
for any $\varepsilon > 0,$  
\begin{align*}
    \lim_{k,N \to \infty} \mathbb{P}(| b^N_k(h^*)-1| > \varepsilon) = 0.
\end{align*}
\end{theorem}

Theorem~\ref{theorem:diagnosability} shows that the Bayesian belief in the true fault \( h^* \in \mathcal{M} \) asymptotically converges to 1 under the conditions 1-4. 
This formalizes diagnosability given the modeled faults, control sequence, moving window length, and noise level. It refines the Convergence Condition in \cite[Pg. 270]{anderson2005optimal}.

However, Theorem~\ref{theorem:diagnosability} does not completely resolve Problem~\ref{prob:PassiveFaultID}. 
Firstly, Problem~\ref{prob:PassiveFaultID} assumes a given control sequence, which is not guaranteed to satisfy strictly positive $\lambda^N(\mathcal{M})$ 
as illustrated by Example~\ref{ex:1}. 
Secondly, if the true fault is not modeled, i.e., \(h^* \notin \mathcal{M}\), the standard Bayesian approach may still assign a high belief to one of the modeled hypotheses, resulting in a misclassification. 
Thirdly, Theorem~\ref{theorem:diagnosability} assumes that all filters remain stable, which may not be satisfied due to model mismatch.
These motivate additional mechanisms to detect unmodeled dynamics and improve fault identification beyond the basic Bayesian framework.

\subsection{Hypothesis Testing for unmodeled faults}

For the analysis with unmodeled fault, where $h^* \notin \mathcal{M}$, we require additional assumptions to identify $\textsc{null}$, i.e., true fault is not in $\mathcal{M}$, for Problem~\ref{prob:PassiveFaultID}. 
For instance, suppose the true fault $h^* \notin \mathcal{M}$ has dynamics ``almost" identical to one of the modeled hypothesis, say $m \in \mathcal{M}$. 
Then it is more reasonable, in practice, to declare that $m$ is identified, rather than \textit{true fault is unmodeled}. 
This motivates the following assumption, which enables a meaningful analysis of failure identification in Problem~\ref{prob:PassiveFaultID}.
\begin{assumption}\label{assump:reasonable_pfid_probelm}
    Suppose the true fault is not modeled $h^* \notin \mathcal{M}$, and the control sequence $\{u_k\}_{k \geq 0}$ is given.
    Then we assume that $h^*$ for Problem~\ref{prob:PassiveFaultID} is diagnosable in the sense of Theorem~\ref{theorem:diagnosability}.
    That is, we assume that the observed information $I_k^N$ subject to stochastic noise is not consistent (Definition~\ref{def:consistent_filter}) to any filter estimates in $\mathcal{M}$. Formally, for all $N \geq 1$
    \begin{align*}
    \bar{\lambda}^N(\mathcal{M}) = \!\! \min_{m \in \mathcal{M}, k\ge N-1} \! &\Big\{ \frac{1}{N} \!\! \sum_{k-(N-1)}^k \!\!\!  \mathbb{E}\big[ \left( y_k - \hat{y}_{m,k|k-1} \right)^T \\
    & S_{m,k}^{-1} \left( y_k - \hat{y}_{m,k|k-1} \right) \big] \Big\},
    \end{align*}
    where $\bar{\lambda}^N > 0$ is asymptotically unbounded.
\end{assumption}

Under this setting, we introduce a hypothesis testing step to reject hypothesis prior to the belief update~\eqref{eq:belief_update}. 
Let \(HT(I_k^N, m; \alpha)\) denote the hypothesis test function at significance level \(\alpha \in (0,1)\). Then, we set that
\begin{equation}
    \label{eq:HT}
    HT(I_k^N, m; \alpha) < 0 \quad \implies \quad p(I_k^N \mid m) = 0.
\end{equation}

In our work, the hypothesis test \(HT(\cdot)\) is a standard two-tailed Chi-square test using the time-average of measurement estimates over the information $I_k^N$ for each $m \in \mathcal{M}$, i.e., for time $k \geq N-1$ (such that $I_k^N \neq \varnothing)$, \eqref{eq:HT} becomes:
\begin{align}
\label{eq:hypo_test}
        &\Bigl( -\bar{\chi}^2 + \frac{1}{N}\textsc{chi2inv}\!\Bigl(\frac{1-\alpha}{2}, n_yN\Bigr) < 0 \Bigr) 
        \; \vee 
        \nonumber \\
        &\Bigl( \bar{\chi}^2 - \frac{1}{N}\textsc{chi2inv}\!\Bigl(\frac{\alpha}{2}, n_yN\Bigr) \! < 0 \Bigr)
    \!\! \implies \!
    p(I_k^N|m) = 0,
\end{align}
where \textsc{chi2inv} is the inverse cumulative distribution function of the chi-squared distribution, and $\bar{\chi}^2 = \frac{1}{N} \sum_{k-(N-1)}^k  \left( e_{m,k} \right)^T S_{m,k}^{-1} \left( e_{m,k} \right)$.

As the available information asymptotically grows ($N \rightarrow \infty$), it is well known that if $m = h^* \in \mathcal{M}$, then $\bar{\chi}^2$ converges to the chi-square distribution via Definition~\ref{def:consistent_filter}. If $m \neq h^*$, then $\bar{\chi}^2$ diverges from the chi-square distribution due to (i) unstable filters, or (ii) $\mathbb{E}[\bar{\chi}^2] \geq \lambda^N(\mathcal{M})$, $\textsc{var}[\bar{\chi}^2]$ is finite, and $\lambda^N(\mathcal{M})$ is asymptotically unbounded as in Assumption~\ref{assump:reasonable_pfid_probelm}.

Intuitively, such hypothesis rejection serves for two purposes. First, if the true fault is modeled ($h^* \in \mathcal{M}$), then the rejection process asymptotically set $p(I_k^N|\tilde{m}) = 0$, where $\tilde{m} \in \mathcal{M}$ denotes the hypotheses with unstable filters. As such, $\tilde{m}$ is ruled out from the likelihood ratio derivation in Theorem~\ref{theorem:diagnosability}. Second, if the true fault is not modeled ($h^* \notin \mathcal{M}$) and Assumption~\ref{assump:reasonable_pfid_probelm} holds, then the rejection process is effective to set all $P(I_k^N|m)$ to zero, avoiding misclassification in the sense of Problem~\ref{alg:pfd}.


By this rejection rule, a scheme to identify that \textit{the true fault is unmodeled} (i.e., \(h^* \notin \mathcal{M}\)) is by checking if all modeled hypotheses are rejected.
However, it is important to note that the significance level \(\alpha\) upper bounds the probability of incorrectly rejecting a true fault. While choosing a lower \(\alpha\) reduces such chance, it increases the risk of failing to reject a false hypothesis. In practice, a finite significance level (e.g., \(\alpha = 0.05\)) is adopted, which implies that there remains a finite risk of incorrectly rejecting the true fault even when \(h^*\) is modeled.

\section{Passive Fault Identification}
\label{sec:passive FID}

In this section, we introduce an algorithm for passive FID in Problem~\ref{prob:PassiveFaultID} based on the framework and analysis in Sec.~\ref{sec:bayes_fid}.
Before presenting our algorithm, we introduce a strategy to mitigate potential misclassifications due to the incorrect rejection of the true fault, to which we refer as the \emph{belief renormalization} procedure.
%
Rather than immediately declaring that the true fault is unmodeled when all modeled hypotheses are rejected by the hypothesis test in~\eqref{eq:hypo_test}, we modify the standard belief update~\eqref{eq:belief_update} as follows:
\begin{equation}\label{eq:renorm}
b_k(\mathcal{M}) =
\begin{cases}
     \frac{1}{|\mathcal{M}|}, & \text{if } \displaystyle\sum_{m \in \mathcal{M}} p(I_k^N \mid m) = 0, \\[1ex]
    \text{ per }~\eqref{eq:belief_update}, & \text{otherwise.}
\end{cases}
\end{equation}
In other words, if the likelihood functions for \emph{all} \(m \in \mathcal{M}\) vanish (i.e., every modeled hypothesis is rejected), the belief is reinitialized as a uniform distribution over \(\mathcal{M}\), representing equal uncertainty rather than assigning a zero probability to each hypothesis.
This renormalization mechanism ensures that transient rejections---arising from either insufficient excitation of the control input or statistical fluctuations---do not immediately lead to misclassification. 
%

\subsection{Passive FID Algorithm}

We present our passive FID algorithm in Algorithm~\ref{alg:pfd}.  
The algorithm takes as input the set of modeled hypotheses \(\mathcal{M}\), the moving window size \(N\), the execution duration \(K \in \mathbb{N}^0\) (determined by the control sequence \(u_0, \ldots, u_K\)), and a user-defined belief threshold \(b_{th} \in \left[\frac{1}{|\mathcal{M}|},1\right]\).  
It returns the identified fault \(m^\textrm{ID} \in \mathcal{M} \cup \{\textsc{null}\}\).  

The algorithm begins by initializing \(b_0\), which can be set as a uniform distribution over \(\mathcal{M}\) or an informed distribution if domain knowledge is available.  
After receiving each measurement \(y_k\), it performs estimation and constructs \(I_k^N\).  
If \(I_k^N\) is not empty (i.e., \(N \geq k-1\)), the algorithm conducts hypothesis testing (Line 5) and updates the belief distribution (Line 6).  
If there exists an \(m \in \mathcal{M}\) such that \(b_k(m) > b_{th}\), then \(m\) is declared as the identified fault.  
Otherwise, if no modeled hypothesis exceeds the threshold throughout the execution duration \(K\), the algorithm returns \textsc{null}.

\begin{algorithm}[t]
\caption{Passive Fault Identification}\label{alg:pfd}
\SetAlgoLined
\SetNlSty{}{}{:}  
\KwIn{\(\mathcal{M}, N, K, \alpha, b_{th}\)
}
\KwOut{Identified fault model \(m^\textrm{ID}\) (or \textsc{null} if undetermined)}
Initialize \(b_0(\mathcal{M})\)\;
\While{\(k \leq K\)}{
    Observe \(y_k\), update filter estimates, and build \(I_k^N\)\;
    \uIf{\(I_k^N \neq \varnothing\)}{
        Perform hypothesis rejection per~\eqref{eq:hypo_test}\;
        Update belief via renormalization per~\eqref{eq:renorm}\;
        \uIf{\(b_k(m) > b_{th}\) for some \(m \in \mathcal{M}\)}{
            \textbf{return} \(m^\textrm{ID}=m\)\;
        }
    }
}
\textbf{return} \(m^\textrm{ID}= \textsc{null} \)\;
\end{algorithm}

\subsection{Convergence of Passive FID Algorithm}

In our passive FID scheme, we first define the failure indicator function
\begin{equation}
F(m^\textrm{ID}) =
\begin{cases}
1 & \text{if } \Bigl( m^\textrm{ID} \neq h^* \text{ and } h^* \in \mathcal{M} \Bigr) \; \lor \\
   & \qquad  \Bigl( m^\textrm{ID} \neq \textsc{null} \text{ and } h^* \notin \mathcal{M} \Bigr), \\[1ex]
0 & \text{otherwise}.
\end{cases}
\end{equation}

We now state the main convergence result of Algorithm~\ref{alg:pfd} for Problem~\ref{prob:PassiveFaultID}.
Let \( \pi^* \in [0,1] \) denote the probability that \( h^* \in \mathcal{M} \). Then the overall failure probability is given by
\begin{equation}
    \mathbb{P}(F) = \pi^* \, \mathbb{P} (F\mid h^* \in \mathcal{M}) + (1-\pi^*)  \mathbb{P} (F\mid h^* \notin \mathcal{M} ).
\label{eq:fail_a}
\end{equation}

Similarly, at each time \( k \ge N-1 \), 
the failure probability can be written as (denoting $A^*$ as the event of rejecting $h^*$)
\begin{align}
    & \mathbb{P}(F)_k = \pi^* \Big( \mathbb{P}(\exists\, m \neq h^*:b_k(m) > b_{th} \mid A^*) \mathbb{P}(A^*) \nonumber \\
    & \quad + \mathbb{P}(\exists\, m \neq h^*: \, b_k(m) > b_{th} \mid \neg A^* ) \mathbb{P}(\neg A^* \Big) \nonumber \\
    & \quad + (1-\pi^*) \, \mathbb{P}(\exists\, m \in \mathcal{M} :\, b_k(m) > b_{th}),
    \label{eq:p_fail_a}
\end{align}
where \( b_{th} \in \left[\frac{1}{|\mathcal{M}|},1\right] \) is a prescribed belief threshold, and the event “reject \( h^* \)” is defined via the hypothesis test in~\eqref{eq:hypo_test}.  

\begin{theorem}[Asymptotic Reliability of Passive FID]\label{theorem:asymp_reliability}
    Let $\mathcal{M}$ and the given control sequence $\{u_k\}_{k \geq 0}$ satisfy the Conditions 1,2,4 in Theorem~\ref{theorem:diagnosability} (without the the stable filter condition).
    Furthermore, if the true fault is not modeled, i.e., $h^* \notin \mathcal{M}$,  Assumption~\ref{assump:reasonable_pfid_probelm} holds.
    Then, Algorithm~\ref{alg:pfd} guarantees that $\mathbb{P}(F)_k \rightarrow 0$ as $N \rightarrow \infty$.
\end{theorem}

Theorem 2 establishes that under consistent filtering (Definition~\ref{def:consistent_filter}), and the diagnosability of unmodeled dynamics (Assumption~\ref{assump:reasonable_pfid_probelm}), the passive fault identification algorithm achieves asymptotic reliability. Specifically:  
\begin{itemize}
    \item If \(h^* \in \mathcal{M}\), the Bayesian belief concentrates on the true fault, overpowering measurement and process noise as \(N \to \infty\).
    \item If \(h^* \notin \mathcal{M}\), statistical inconsistency ensures all modeled hypotheses are rejected.  
\end{itemize}

In practice, Theorem 2 guarantees that for any desired misidentification tolerance \(\delta \in (0,1) \), there exists a finite moving window \(N(\delta)\) such that \(\mathbb{P}(F)_k \leq \delta\).
 
While the passive method ensures asymptotic convergence, its performance depends critically on the \emph{given} control sequence’s ability to excite hypothesis-specific dynamics. In practice, arbitrary control inputs may poorly regulate the signal-to-noise ratio \(\lambda^N\), leading to slow convergence or even impossible fault identification as illustrated in Example~\ref{ex:1}.
This motivates \emph{active fault identification}, where control sequences are strategically designed to accelerate the decay rate of \(\mathbb{P}(F)_k\) by maximizing a proxy of \(\lambda^N\).

\section{Active Fault Identification}
\label{sec:active FID}

Our active FID paradigm that addresses Problem~\ref{prob:ActiveFaultID} designs control inputs \( u_k \) to maximize the discriminative power between hypotheses. 
Motivated by Theorem~\ref{theorem:diagnosability}, we aim to maximize the smallest distance between the predicted measurement estimates of the true ($\hat{y}_{h^*,k|k-1},\;h^* \in \mathcal{M}$) and the others ( $\hat{y}_{m,k|k-1}, S_{m,k},\;m \in \mathcal{M} \setminus \{h^*\}$ ). 
However, this is challenging because $h^*$ is unknown a priori.
A reasonable proxy is the arithmetic mean of the pairwise distance between \emph{any} hypotheses.
Nevertheless, this measure is not appropriate because it may lead to a large distance between one pair of hypotheses, while sacrificing the separation of other pairs that may include the true fault $h^*$. 

As such, we propose to use geometric mean as the measure. Formally, the optimization problem for active fault identification is:
\begin{subequations}
    \label{eq:opt}
    \begin{align}
        & u_{k}^* = \argmax_{u_k \in U_a} J(u_k), \label{eq:opt:a}\\
        & J(u_k) := \Big(\!\! \prod_{m \neq m' \in \mathcal{M}} \!\!\!d\big(f_{m,k}(u_k), f_{m',k}(u_k)\big)\Big)^{|\mathcal{M}|^{-2}}\!\!\!, \label{eq:opt:b}\\
        & f_{m,k}(u_k) := \mathcal{N}(\hat{y}_{k|k-1},S_{k} \mid m, u_k), \label{eq:opt:c} \\
        & d(f_{m}, f_{m'}) = (\hat{y}_{m} - \hat{y}_{m'})^T S_{m'}^{-1} (\hat{y}_{m} - \hat{y}_{m'}),\label{eq:opt:d}
    \end{align}
\end{subequations}
where \( U_a \subseteq U \) is a compact admissible control set, 
$f_{m,k}(u_k)$ is the predicted next-step measurement distributions using control $u_k$ under hypothesis $m$,
\( d(\cdot,\cdot) \) is the Mahalanobis distance, and \( J(u_k) \) represents the \emph{geometric average} of the pairwise hypothesis separation.

To analyze the property of the optimal solution to~\eqref{eq:opt}, we characterize the admissible control set $U_a$.
\begin{definition}[Degenerate Admissible Control]\label{def:degenerate}
The admissible control set $U_a$ is called degenerate for~\eqref{eq:opt} if \( J(u) = J(u') \) for all \( u, u' \in U_a \). Otherwise, it is \emph{non-degenerate}.
\end{definition}
\begin{lemma}\label{lemma:active}
    Let $\{u_k^*\}_{k \geq 0}$ be a control sequenced where each $u_k^*$ is the optimal solution to~\eqref{eq:belief_update} over non-degenerate admissible control set $U_a$.
    Suppose $h^* \in \mathcal{M}$. Then, $\{u_k^*\}_{k \geq 0}$ guarantees that 
    diagnosibility of $\mathcal{M}$ is not fundamentally limited, i.e.,
    \[
        \lambda^N(\mathcal{M}) > 0 \qquad \forall m \in \mathcal{M} \setminus \{h^*\}.
    \]
\end{lemma}

By the construction of~\eqref{eq:opt}, the active control seeks to maximize the \emph{averaged} pairwise distance at each time $k$. 
In addition, by Lemma~\ref{lemma:active}, if the admissible control set $U_a$ is non-degenerate (i.e., rich enough), then the active control guarantees strictly positive $\lambda^N(\mathcal{M})$, satisfying 
the necessary condition for reliable FID via Bayesian belief update (but not the sufficient condition for asymptotic fault diagnosability in Theorem 1).


The active fault identification algorithm follows the same steps as in Algorithm~\ref{alg:pfd} with the addition of the following after Line 8:
\begin{equation*}
    \text{\footnotesize Line 9:} \;\; \text{ Compute }  u_k^* = \argmax_{u \in U_a} J(u) \;\; \text{(per~\eqref{eq:opt})}
\end{equation*}
The enhanced performance of active FID is demonstrated via  numerical experiments in Sec.~\ref{sec:exp}.

\section{Numerical Experiments}\label{sec:exp}
We present two case studies to demonstrate the convergence and performance of Passive vs.\ Active FID: a nonlinear \textit{two-water tank system}~\cite{paulson2014guaranteed} and a \textit{Mars satellite system} with discontinuous attitude dynamics due to MRP switching~\cite{schaub2003analytical}. 
The baseline rejection confidence and belief identification threshold are set to $\alpha=0.05, b_{th}=0.95$, respectively.
The code is available on GitHub~\cite{FID:github}.
Due to the page limit, we refer the reader to this GitHub page for full descriptions of each experiment.

\subsection{Case Study 1: Two-Water Tank System}
\begin{figure}[t]
    \centering
    \begin{subfigure}[b]{0.51\linewidth}
        \centering
        \includegraphics[width=\linewidth]{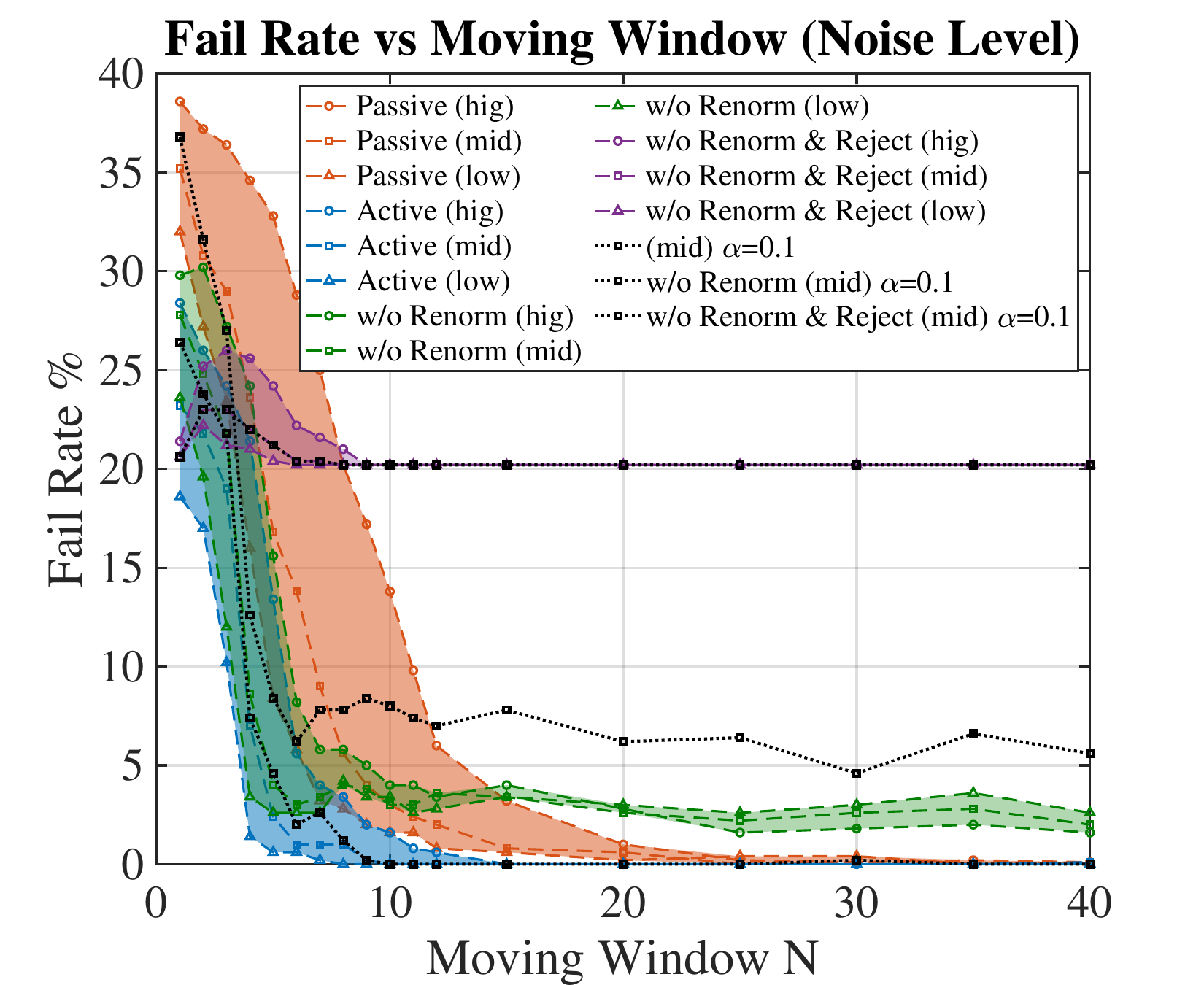}
        \caption{failure ID rate}
        \label{fig:subfig1}
    \end{subfigure}
    \hspace{0em}
    \begin{subfigure}[b]{0.45\linewidth}
        \centering
        \includegraphics[width=\linewidth]{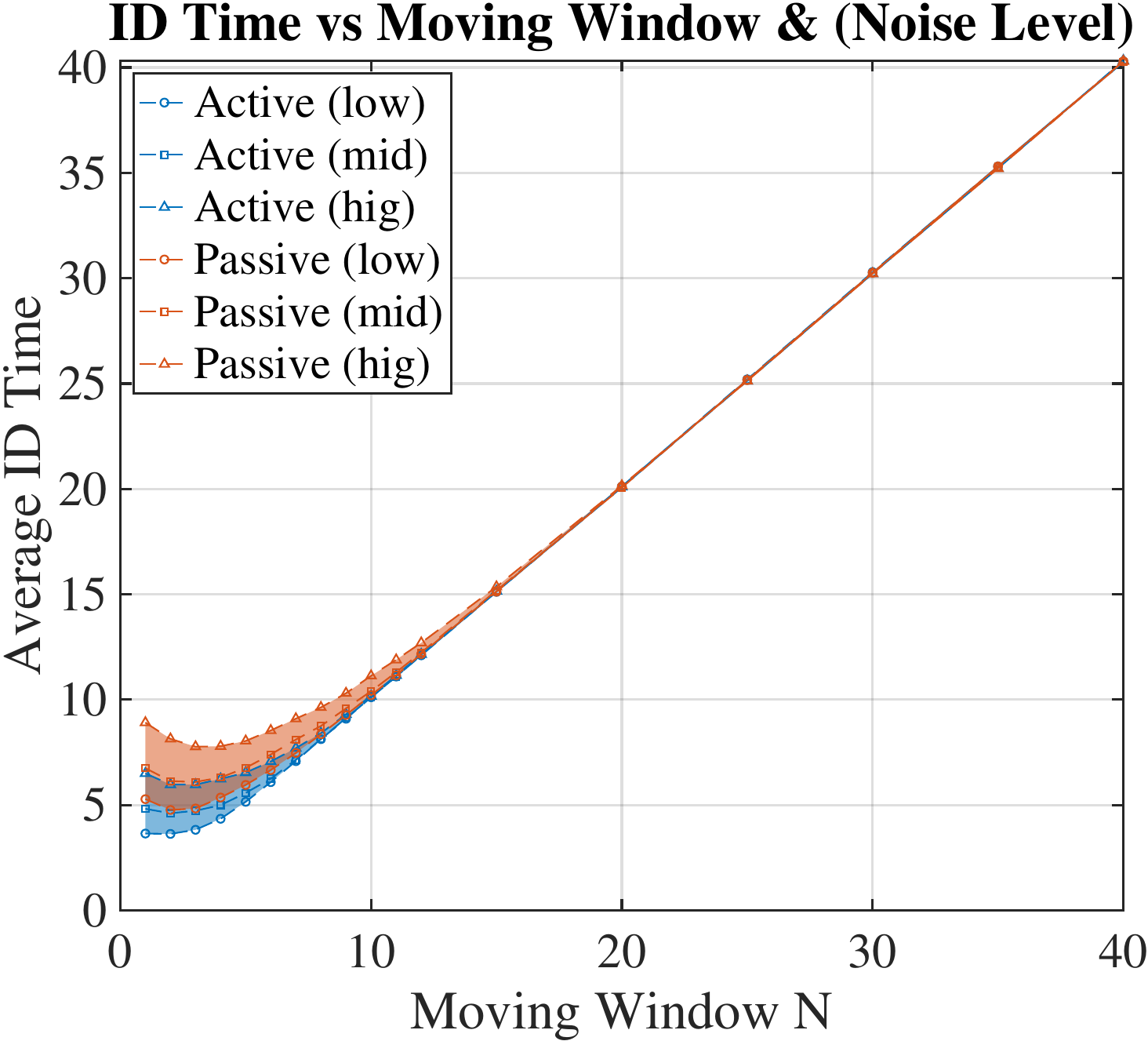}
        \caption{average ID time}
        \label{fig:subfig2}
    \end{subfigure}
    \caption{Active vs Passive FID performance on Case: 1.
    }
    \label{fig:watertank}
\end{figure}

Adapted from~\cite{paulson2014guaranteed}, this case study considers three types of fault: leakage of tank 1, leakage of tank 1 to tank 2, and leakage of tank 2. 
Fig.~\ref{fig:watertank} shows two performance metrics: (a) failure identification rate in percentage and (b) average identification time (delay). 
\emph{Each point} is a test configuration evaluated over 500 Monte Carlo trials with randomized hypotheses (probability of the true fault being modeled: $\pi^*=0.8)$, initial states, and process noise.
Different measurement noise levels are tested to show robustness.
Fig.~\ref{fig:subfig1} shows Active FID achieves near zero failure rate as $N$ increases, whereas Passive FID reduces the failure rate more slowly. 
Fig.~\ref{fig:subfig2} shows comparable identification delays between the two methods.

\paragraph*{Ablation Study}  
We study the impact of algorithm components on asymptotic reliability (Fig.~\ref{fig:subfig1}). 
Without renormalization, the failure rate remains finite, which is upper bounded by~\eqref{eq:p_fail_a}: $\pi^* \alpha = 4\%$. Removing both renormalization and rejection causes the failure rate to converge to $1-\pi^* = 20\%$, confirming theoretical predictions in~\eqref{eq:p_fail_a}.
The black dotted lines in Fig.~\ref{fig:subfig1} validate results with a larger rejection parameter ($\alpha=0.1$).
Similarly, we observe that failure rates drop to zero with full components, are bounded by $\pi^* 0.1=8\%$ without renormalization, and converge to $1-\pi^*=20\%$ without both renormalization and rejection.

\subsection{Case Study 2: Mars Satellite}

\begin{figure}[t]
    \centering
    \begin{subfigure}[b]{0.51\linewidth}
        \centering
        \includegraphics[width=\linewidth]{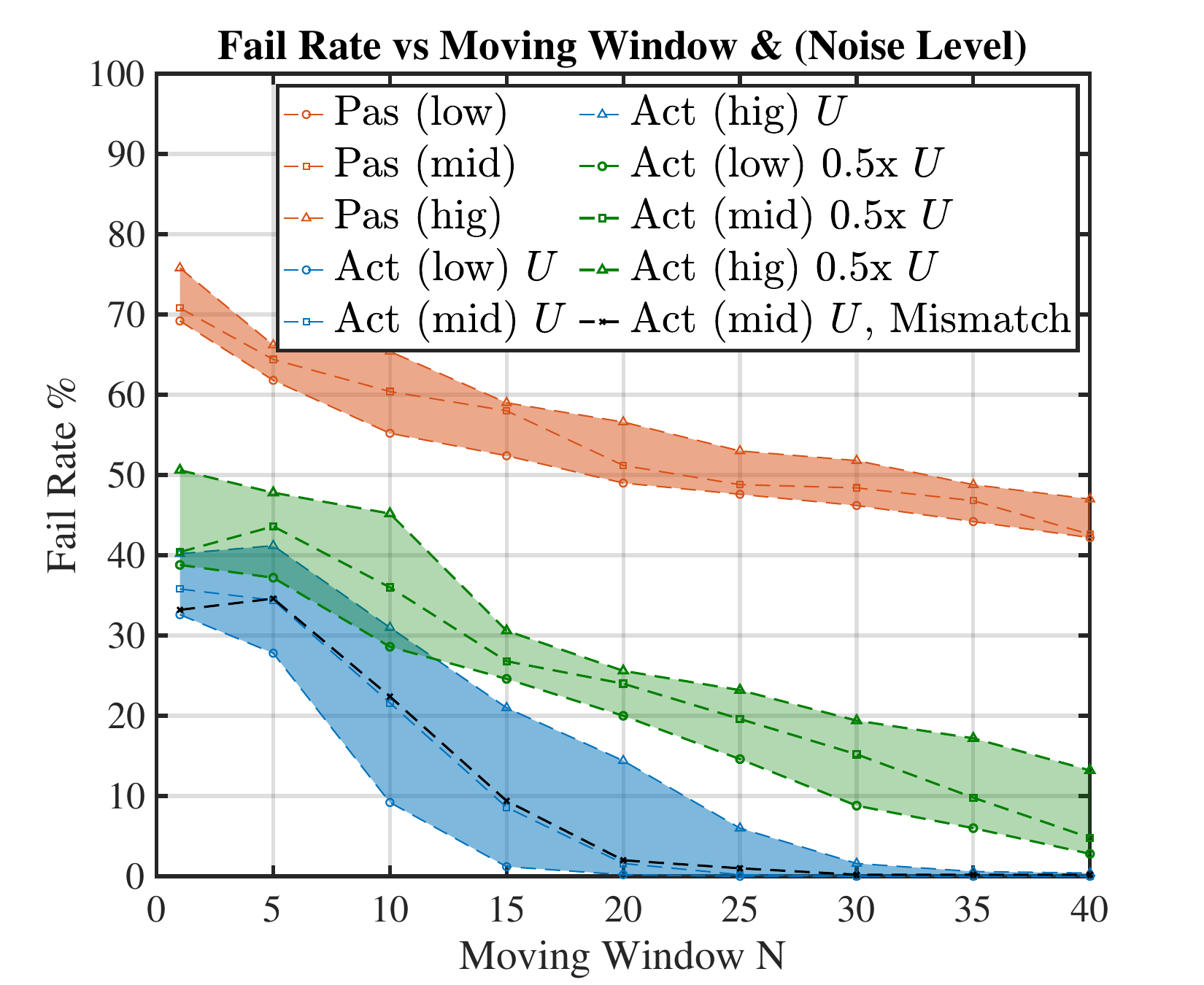}
        \caption{failure ID rate}
        \label{fig:mars_sat_a}
    \end{subfigure}
    \hspace{0em}
    \begin{subfigure}[b]{0.45\linewidth}
        \centering
        \includegraphics[width=\linewidth]{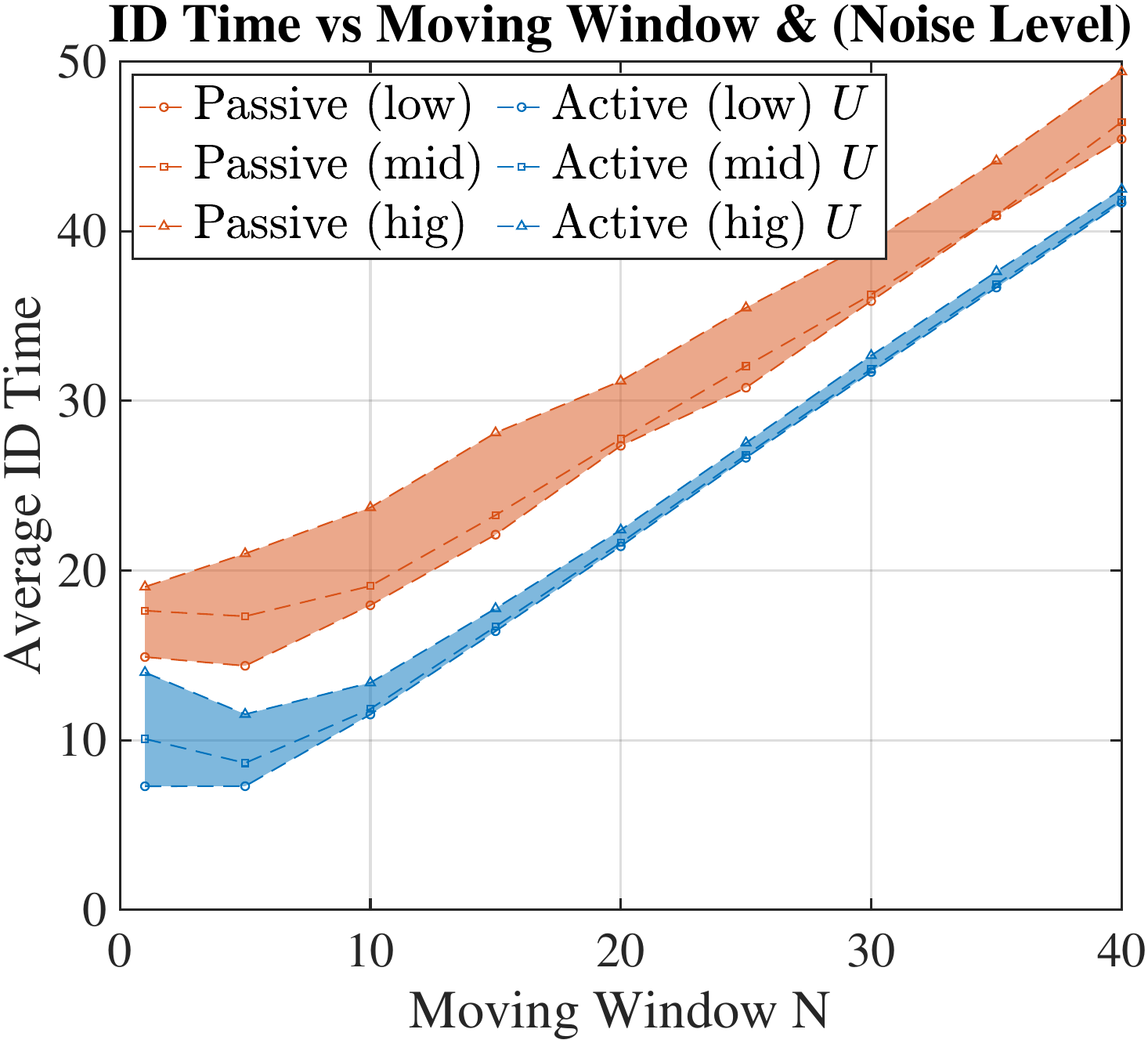}
        \caption{average ID time}
        \label{fig:mars_sat_b}
    \end{subfigure}
    \caption{Active vs Passive FID performance on Case: 2.
    }
    \label{fig:mars_sat}
\end{figure}

In the Mars satellite case (Fig.~\ref{fig:mars_sat}), \emph{each point} represents 500 Monte Carlo trials with randomized hypotheses, initial states, and process noise. 
Faults are modeled as reduced torques along the three principal axes (1–3), with 0 denoting the nominal system.
Passive FID, using a control law designed for attitude tracking, struggles with failure rates above 45\% even at large 
moving window $N$.
In contrast, Active FID rapidly reduces failures. For both methods, identification delay grows linearly with 
$N$ as $I_k^N$requires at least $N$ steps, but Active FID achieves noticeably shorter delays.

\paragraph*{Control Authority Effect}  
We study control authority by comparing full control ($U_a = U$) with reduced authority ($U_a = \{0.5 u \mid u \in U\}$). 
As expected, limited authority degrades performance, highlighting the advantage of fully exploiting control freedom in Active FID.

\paragraph*{Model Mismatch}
We study the effect of modeling mismatch on Active FID performance.
Modest discrepancies---such as slight differences in fault parameters or unknown disturbance---have little impact, as shown by the black dashed line in Fig.~\ref{fig:mars_sat_a}.
For illustration, we test random 10\% deviations in fault parameters and disturbance torques up to 10\% of the maximum controlled torque.

\begin{figure}[htbp]
    \centering
    \begin{subfigure}[b]{0.8\linewidth}
        \centering
        \includegraphics[width=\linewidth]{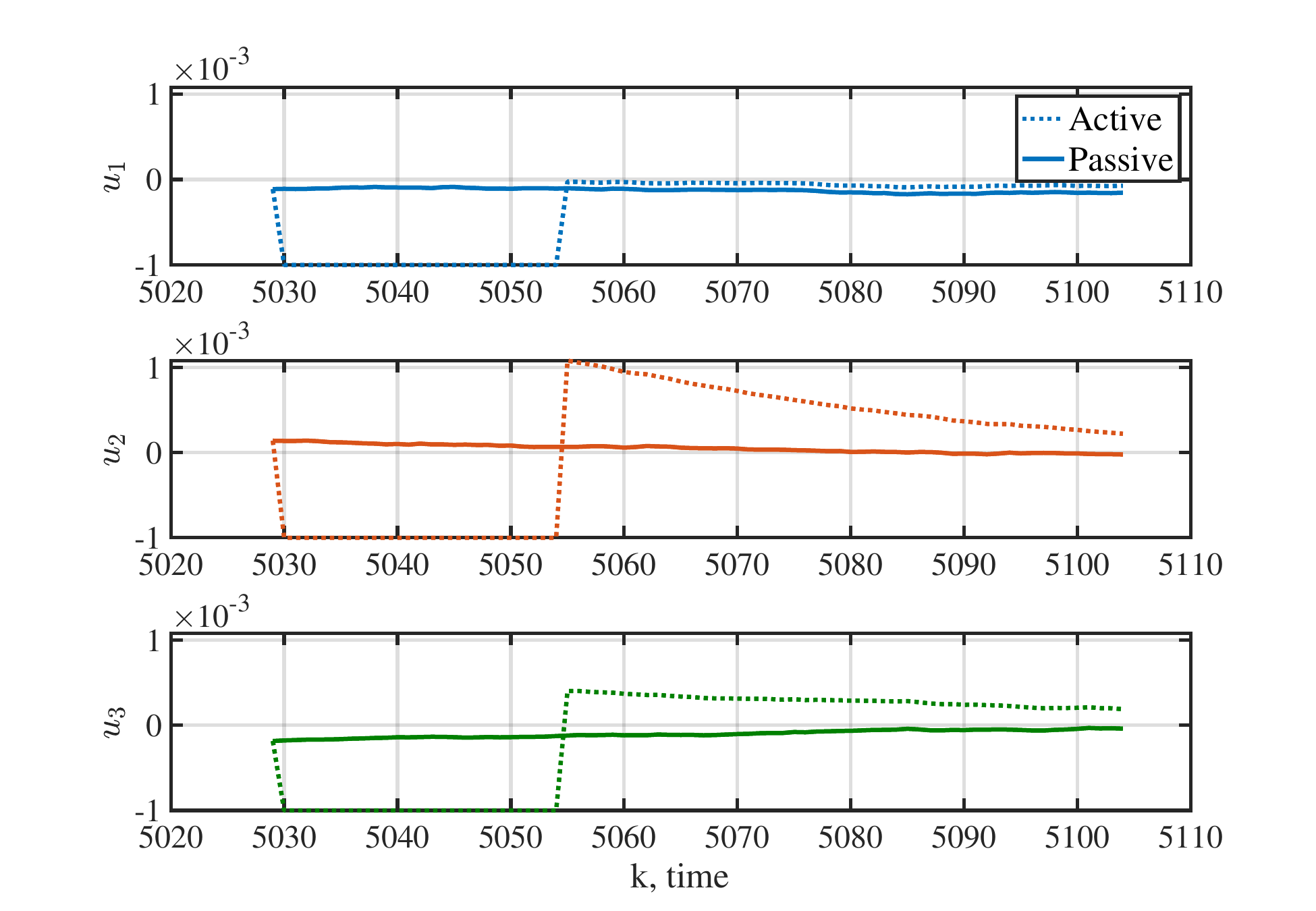}
        \caption{control trajectories
        }
    \end{subfigure}
    
    \begin{subfigure}[b]{0.8\linewidth}
        \centering
        \includegraphics[width=\linewidth]{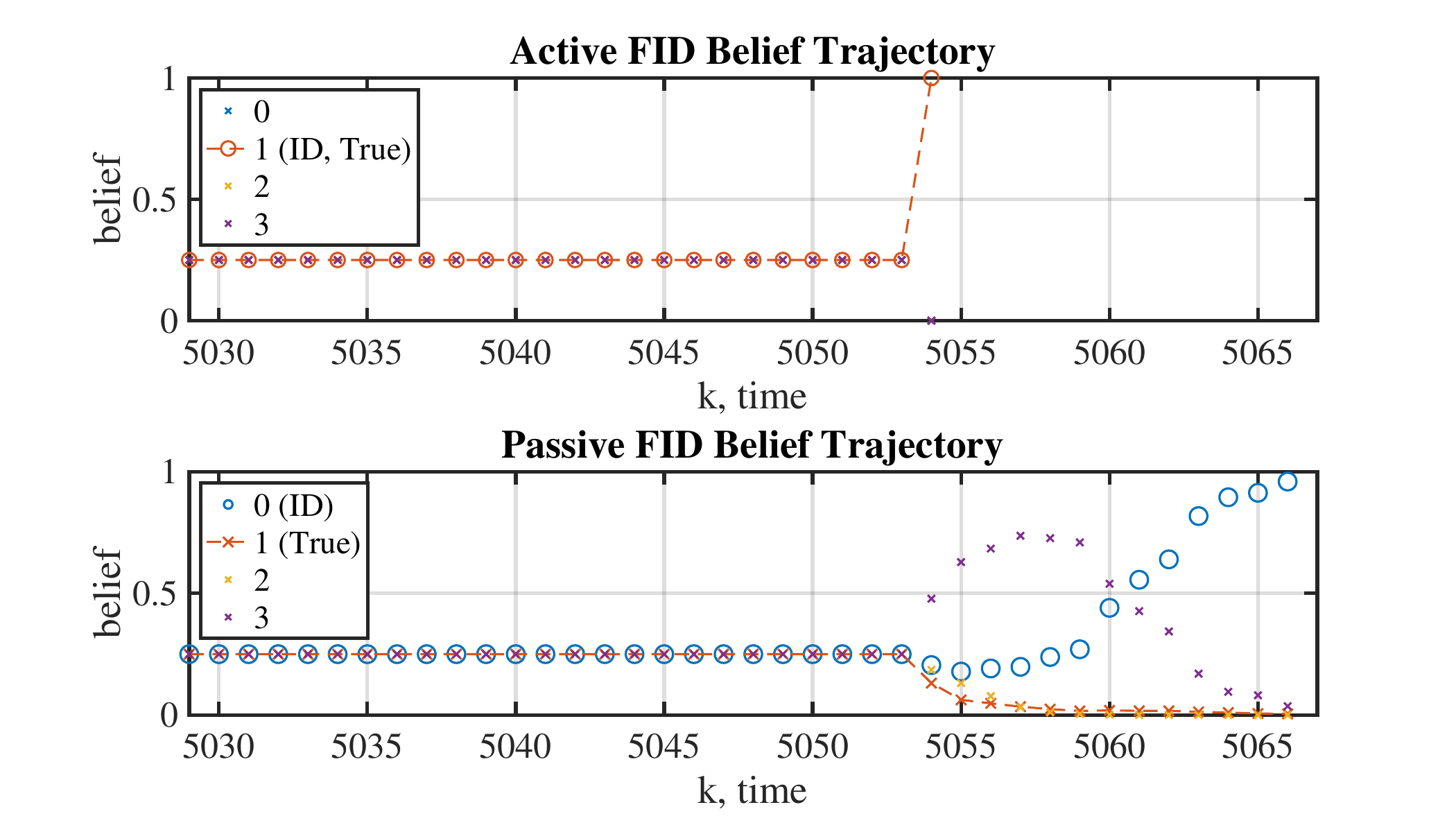}
        \caption{belief trajectories}
    \end{subfigure}
    
    \vspace{1em} 
    
    \begin{subfigure}[b]{0.7\linewidth}
        \centering
        \includegraphics[width=\linewidth]{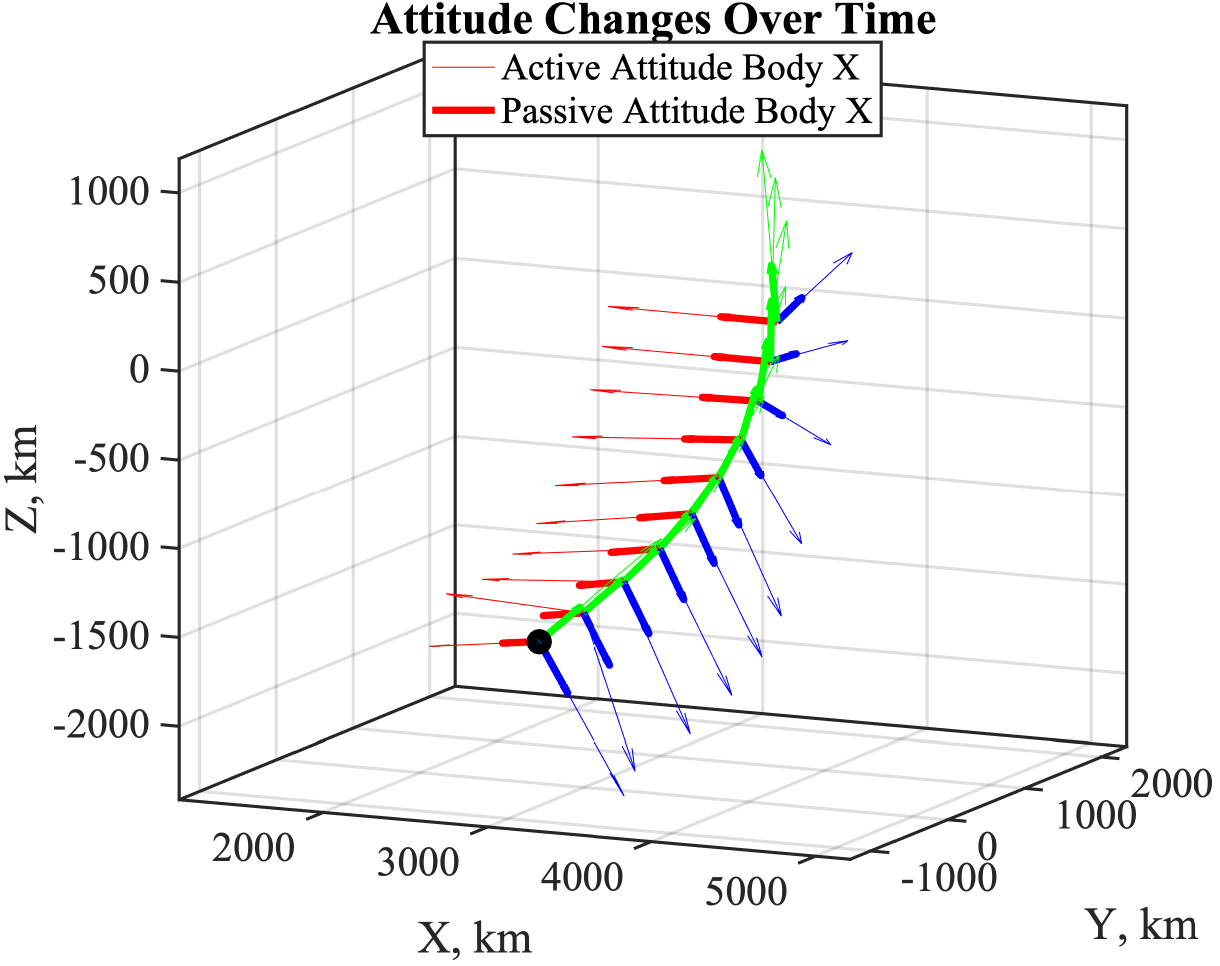}
        \caption{attitude trajectories}
    \end{subfigure}
    \caption{Control, hypothesis belief, attitude trajectories of Active vs Passive FID of a random trial, with 25 moving window and high measurement noise.}
    \label{fig:mars_sat_trial3}
\end{figure}

\paragraph*{Trajectory Comparison of Single Trial}
Fig.~\ref{fig:mars_sat_trial3} compares Active and Passive FID on a single Mars satellite trial (same random seed, $N=25$, high noise). 
Active FID correctly identifies the true fault: 1, while Passive FID fails to distinguish between hypotheses 0 and 1. 
Despite this, attitude trajectories are similar due to limited active control magnitude and its short duration. 
A video showing the full trajectory evolution is available at the project website~\cite{FID:github}.

In summary, our experiments show that Active FID achieves lower failure rates and shorter identification delays than Passive FID. 
The proposed algorithms are asymptotically reliable under varying noise levels, and their performance improves as control authority increases.
Moreover, small-scale tests demonstrate robustness to parameter variations and modest modeling mismatches.
%


\section{Conclusion}
\label{sec:conclusion}
We introduced a new quantitative diagnosticability definition for discrete-time nonlinear systems subject to additive Gaussian white noise. 
Building on this definition, we developed a Bayesian fault identification framework capable of handling both modeled and unmodeled faults. 
The empirical studies showed that the risk of failure identification for both Passive and Active FID asymptotically decreases as more information is available; the Active FID outperforms the Passive FID in correctly identify true fault with shorter time.
In the future work, we plan to extend this framework for incipient faults (developing gradually) by augmenting the hypothesized fault models with dynamic fault parameters.

\section{Proofs}
\label{sec:proofs}
{

\subsection{Proof of Theorem~\ref{theorem:diagnosability}}
\begin{proof}[Proof]
\textbf{Step 1: Log-Likelihood Ratio.}  
The likelihood is
\(
p(I_k^N|m)=\prod_{k-N+1}^{k} f\bigl(y_i-\hat{y}_{m,k|k-1};0,S_{m,k}\bigr),
\)
and for any $m\neq h^*$ we define the log-likelihood ratio
\[
\begin{split}
& L_N^{m,h^*} := \ln\frac{p(I_k^N|m)}{p(I_k^N|h^*)} = \frac{1}{2}\sum_{k-N+1}^{k} 
\Bigl[ \\ &
           - e_{m,k}^T S_{m,k}^{-1} e_{m,k}+ \ln\bigl(\frac{\det S_{h^*,k}}{\det S_{m,k}}\bigr)
           + e_{h^*,k}^T S_{h^*,k}^{-1} e_{h^*,k} \Bigr].
\end{split}
\]
\textbf{Step 2: Innovation Decomposition and Quadratic Expansion.}  
Write
\(
e_{m,k}=y_k-\hat{y}_{m,k|k-1}=(\hat{y}_{h^*,k|k-1}-\hat{y}_{m,k|k-1})+e_{h^*,k},
\)
where $e_{h^*,k}=y_k-\hat{y}_{h^*,k|k-1}$.
Then 
\(
e_{m,k}^T S_{m,k}^{-1}e_{m,k}=A_k+B_k+C_k,
\)
where
\(
A_k=(\hat{y}_{h^*,k|k-1}-\hat{y}_{m,k|k-1})^T S_{m,k}^{-1} (\hat{y}_{h^*,k|k-1}-\hat{y}_{m,k|k-1}),
B_k=2(\hat{y}_{h^*,k|k-1}-\hat{y}_{m,k|k-1})^T S_{m,k}^{-1} e_{h^*,k},
C_k=e_{h^*,k}^T S_{m,k}^{-1}e_{h^*,k}.
\)
Thus,
\[
\begin{split}
L_N^{m,h^*} = \frac{1}{2} \sum_{k-N+1}^{k} \Biggl[ & -\Bigl(A_k+B_k+C_k\Bigr) \\
& +\, e_{h^*,k}^T S_{h^*,k}^{-1}e_{h^*,k} +\, \ln\frac{\det S_{h^*,k}}{\det S_{m,k}} \Biggr].
\end{split}
\]
\textbf{Step 3: Expectation.}  
By the assumption of stable filters, $S_{k} \rightarrow S_{ss}$ as $k \rightarrow \infty$ (for $k \geq N-1$). In addition, $\operatorname{Tr}(S_{m,k}^{-1}S_{h^*,k}) > 0$, and the determinants exist because the innovation covariance is positive definite. Using $\mathbb{E}[B_i]=0$ and $\mathbb{E}[C_i]=\operatorname{Tr}(S_{m,i}^{-1}S_{h^*,i})$, we obtain
\[
\begin{split}
\mathbb{E}[L_N^{m,h^*}] = \frac{1}{2}\sum_{k-N+1}^{k} \Bigl[\, & -\mathbb{E}[A_k] - \operatorname{Tr}\bigl(S_{m,k}^{-1}S_{h^*,k}\bigr) \\
& +\, \operatorname{Tr}(I_{n_y}) + \mathbb{E}\Bigl[\ln\frac{\det S_{h^*,k}}{\det S_{m,k}}\Bigr] \Bigr].
\end{split}
\]
Since $\sum_{i}\mathbb{E}[A_i]\ge N\,\lambda^N(\mathcal{M})$, it follows that
\[
\mathbb{E}[L_N^{m,h^*}]\le \frac{1}{2}\Bigl(-N\,\lambda^N(\mathcal{M})+\mathcal{O}(N)\Bigr),\; \text{as } N \rightarrow \infty.
\]
\textbf{Step 4: Asymptotic Convergence.}  
Assume $\lambda^N(\mathcal{M})\sim\ln N$ and $Var(L_N^{m,h^*})=\mathcal{O}(N)$. Define $\widetilde{L}_N=L_N^{m,h^*}/N$, so that
\[
Var(\widetilde{L}_N)=\mathcal{O}(1/N)\to 0.
\]
Then, by Chebyshev’s inequality, for any $\delta\in(0,1)$,
\[
\mathbb{P}\Bigl(\widetilde{L}_N\ge -\delta\,\lambda^N(\mathcal{M})\Bigr)\to 0,
\]
or equivalently,
\[
\mathbb{P}\Bigl(L_N^{m,h^*}\ge -\delta\,N\,\lambda^N(\mathcal{M})\Bigr)\to 0.
\]
Since
$b_N(m)\propto\exp\bigl(L_N^{m,h^*}\bigr),$
for any $\epsilon>0$ we have $\mathbb{P}\bigl(b_N(m)\ge\epsilon\bigr)\to 0$, and hence
\[
b_N(h^*)=1-\sum_{m\neq h^*}b_N(m)\xrightarrow{p}1.
\]
\end{proof}

\subsection{Proof of Theorem~\ref{theorem:asymp_reliability}}
\begin{proof}
Suppose $h^* \in \mathcal{M}$.
By the consistent hypothesis test, the set of hypotheses $\tilde{M}  \subseteq \mathcal{M} \setminus \{h^* \}$ corresponding to diverged filter estimates is rejected as $N \rightarrow \infty$. Hence, Theorem~\ref{theorem:diagnosability} still holds to distinguish between true fault $h^*$ and false hypotheses in $\mathcal{M} \setminus( \{h\}^* \cup \tilde{M})$ as $N \rightarrow \infty$.

Under the condition of falsely rejecting $h^*$, $b_k(m) = \frac{1}{|\mathcal{M}|}$ for all $m \in \mathcal{M}$ because (i) all hypotheses are rejected---$h^*$ is rejected, and all other $m \in \mathcal{M} \setminus \{h\}^*$ are rejected because they either correspond to unstable filters or statistically diagnosable (per Assumption~\ref{assump:reasonable_pfid_probelm}) from the true whitened statistic, then (ii) renormalization occurs. Therefore
\[ 
\mathbb{P}\Bigl(\exists\, m \neq h^*: \, b_k(m) > b_{th} \,\Big|\, \text{reject }h^*\Bigr) \rightarrow 0,\; \text{as } N \rightarrow \infty.
\]
Secondly, under the condition of not rejecting $h^*$, the belief concentrates on $b_k(h^*)$, hence,
\[ 
\mathbb{P}\Bigl(\exists\, m \neq h^*: \, b_k(m) > b_{th} \,\Big|\, \neg \text{reject }h^*\Bigr) \rightarrow 0,\; \text{as } N\rightarrow \infty.
\]

Suppose \( h^* \notin \mathcal{M} \). Since we assume such $h^*$ generates information $I_k^n$ inconsistent with any filters in $\mathcal{M}$, all the modeled hypothesis $m \in \mathcal{M}$ are rejected by consistent hypothesis test as \( N \to \infty \). 
Specifically,
\[
\mathbb{P}\Bigl(\exists\, m \in \mathcal{M} :\, b_k(m) > b_{th}\Bigr) \le 1 - \mathbb{P}\Bigl(\text{reject all } m \in \mathcal{M}\Bigr) \to 0.
\]
Combining these results in~\eqref{eq:p_fail_a} and the fact that $P(\text{reject } h^*) \leq \alpha$ (denoting $A^*$ as the event of rejecting $h^*$), we see that as $N \rightarrow \infty$,
\[
\begin{split}
& \mathbb{P}(F)_k = \pi^*\Big[\mathbb{P}\Bigl(\exists\,m\neq h^*:b_k(m)>b_{th}\mid A^*)\,\mathbb{P}\Bigl(A^*\Bigr) \\
& \quad + \mathbb{P}\Bigl(\exists\, m \neq h^*: \, b_k(m) > b_{th} \,\Big|\, \neg A^*\Bigr) \,\mathbb{P}\Bigl(\neg A^*\bigr) \Biggr] \\
& \quad + (1-\pi^*) \, \mathbb{P}\Bigl(\exists\, m \in \mathcal{M} :\, b_k(m) > b_{th}\Bigr) \\
& \rightarrow \pi^*\Big[0 \cdot \alpha + 0 \cdot (1-\alpha) \Big] + (1-\pi^*) \cdot 0 = 0.
\end{split}
\]
\end{proof}

\subsection{Proof of Lemma~\ref{lemma:active}}
\begin{proof}
    Let $\{u_k^*\}_{k \geq 0}$ be a control sequenced where each $u_k^*$ is the optimal solution to~\eqref{eq:belief_update} over non-degenerate admissible control set $U_a$. Then $J(u_k^*) > 0$ for all $k$, because if the optimal cost is zero, it implies $J(v_k) = 0$ for all $v_k \in U_a$, which contradicts the non-degenerate $U_a$. Since $J(\cdot)$ is the product of each pairwise Mahaldistance~\eqref{eq:opt:d}, it implies for any $m_i \neq m_j \in \mathcal{M}$:
    \[
    (\hat{y}_{m_i,k|k-1} - \hat{y}_{m_j,k|k-1})^T S_{m_j,k}^{-1} (\hat{y}_{m_i,k|k-1} - \hat{y}_{m_j,k|k-1}) > 0
    \]
    for all $k$. Take $m_i = h^*$, and by the definition of $\lambda^N(\mathcal{M})$ in Definition~\ref{def:diagnos}, the proof is complete.
\end{proof}

}


\bibliographystyle{unsrt}
\bibliography{reference}


\end{document}